 \documentclass[final,5p,times,twocolumn,authoryear]{elsarticle}

\usepackage{amssymb}
\usepackage{lipsum}
\usepackage{amsmath}

\journal{New Astronomy}

\begin{document}

\begin{frontmatter}



\title{A generalized static spherically symmetric anisotropic compact star model in isotropic coordinates}


\author[first]{B. S. Ratanpal}
\affiliation[first]{organization={Department of Applied Mathematics, Faculty of Technology \& Engineering, The Maharaja Sayajirao University of Baroda},
            addressline={}, 
            city={Vadodara},
            postcode={390001}, 
            state={Gujarat},
            country={India}}
\author[first]{Bhavesh Suthar}

\begin{abstract}
In this article, an exact solution of Einstein's field equations for spherically symmetric anisotropic matter distributions in isotropic coordinates is obtained. For this, the solution has been obtained by using a generalized physically significant metric potential $B(r)$ and a specific choice of the anisotropy. This new class of generalized solutions is singularity-free, which can be used to describe relativistic compact stars.
\end{abstract}




\end{frontmatter}




\section{Introduction}
\label{introduction}

An investigation of the solution to the Einstein's field equations reveals that the exact solution is essential for the development of many areas of the gravitational field, including black hole solutions, solar system tests, gravitational collapse and others. In astronomy, the word 'compact object' refers to white dwarfs, neutron stars and black holes that arise as a result of slow gravitational collapse. We all know that stars are isolated bodies, constrained by self-gravity and radiating energy from an internal source. According to the strange matter hypothesis, strange quark matter could possibly be more stable than nuclear matter, and so neutron stars should be mostly made up of pure quark matter. In the context of relativistic astrophysics, the singularity-free interior solutions of the compact object have significant consequences. It is well-known that the model of a compact star can be generated by solving the Einstein's field equations in the framework of general relativity.

\cite{SK1916} discovered the first solution to the Einstein's field equations describing a self-gravitating bounded object nearly a century ago. The Schwarzschild interior solution explains a sphere with uniform density. It was the first approximation to describe the gravitational field of a static spherically symmetric object. Numerous exact solutions to Einstein's field equations are available in the literature; however, not all of them are physically relevant. Out of the 127 solutions identified by \cite{Delgaty1998}, only 16 meet the criteria for describing a physically plausible system. There are two general categories of solutions that describe relativistic stars: isotropic and anisotropic. When the densities of compact stars exceed the density of nuclear matter, unequal principal stresses are expected, known as the anisotropic effect. \cite{Jeans1922} first predicted the presence of anisotropy for self-gravitating objects in the Newtonian regime. Later, \cite{Lemaitre1933} investigated how anisotropy affects objects in the context of general relativity, demonstrating that anisotropy can change the maximum attainable value of the surface gravitational potential. In the theoretical study of relativistic stars \cite{Ruderman1972} that the pressure inside highly dense astrophysical objects, such as X-ray pulsars and certain types of neutron stars with core densities greater than the nuclear density ($10^{15}g.cm^{-3}$), is anisotropic, i.e., the pressure inside these compact objects can be divided into two parts: the radial pressure pr and the tangential pressure pt. According to \cite{Kipp1990}, anisotropy in relativistic stars may be caused by the presence of a solid core or type 3A superfluid. \cite{Herrera1997} investigated the local anisotropy of a self-gravitating system in both Newtonian and general relativistic contexts. According to \cite{Weber1999}, high magnetic fields can also cause anisotropic pressure within a compact object. Anisotropy originates from various reasons, including exotic phase transitions during gravitational collapse, pion condensation \cite{Sokolov1980}, \cite{Herrera1989}, slow rotation of fluid \cite{Herrera1995}, viscosity \cite{Ivanov2010}, and other physical phenomena. \cite{DG2002}, \cite{DG2003}, and \cite{GD2004} also discussed an anisotropic star model utilizing a particular form of matter density. Since then, researchers have conducted numerous studies to discover new exact solutions with anisotropic matter distribution [\cite{SP2016}, \cite{Govender2016}, \cite{Maurya2017}, \cite{Thomas2017}, \cite{Jaya2020}, \cite{Bhar2021}, \cite{Ratanpal2023}, \cite{Mathias2023}, \cite{Das2024}, \cite{GD2024}, and \cite{Kumar2024}]. The results in the references demonstrate that pressure anisotropy is important in the modeling of stellar objects.

Two traditional approaches are usually adopted to solve the Einstein's field equations, which describe physically realistic stellar models. In the first approach, a particular equation of state can be used as a starting point, and the integration begins at the star's core with a specified central pressure. This integration is iterated until the pressure vanishes, indicating that the star's surface has been reached. The second method involves solving Einstein's field equations. It is quite challenging to find an exact solution to an underdetermined system of nonlinear second-order ordinary differential equations. In the case of a static anisotropic matter distribution, the Einstein's field equations can be reduced to three ordinary differential equations with five unknowns. To obtain exact solutions, one can solve the Einstein's field equations by making assumptions about gravitational potentials, energy density, or anisotropy. Nevertheless, it would be preferable to have a general method that generates exact solutions in a systematic manner. Some systematic techniques developed in the past include those of \cite{SC2006}, \cite{RV2002}, \cite{Lake2003}, \cite{DM2004}, \cite{Herrera2004}, and \cite{CM2006}. General relativity allows us to employ any well-defined coordinate system. The references listed above mostly make use of the canonical coordinate system. In certain circumstances, the transformation to Schwarzschild coordinates is not well defined. This phenomenon can be seen when isotropic coordinates are used. \cite{Nariai1950} demonstrated that every line element provided in Schwarzschild-like coordinates may be transformed into an isotropic one, but the opposite is not always feasible. This appears to illustrate that isotropic coordinates are more general than canonical coordinates. All three spatial dimensions are treated equally in the isotropic coordinate system, which is its main feature. The use of an isotropic coordinate system may reveal new insights that lead to new solutions. Numerous researchers have done significant work to develop an anisotropic compact star model in isotropic coordinates [\cite{Govender2015}, \cite{N2015}, \cite{Pant2015}, \cite{Bhar2018}, \cite{Heras2019}, and \cite{Suthar2024}].

Motivated by aforementioned works, in this article, an exact solution to Einstein's field equations in an isotropic coordinate system is obtained by adopting a generalized form of anisotropy. Our solution, which describes the interior of the star, is found to be continuous and non-singular. This article is structured as follows: In Section \ref{sec2}, the Einstein's field equations corresponding to the anisotropic compact object have been discussed. Section \ref{sec3} derives the generating function and corresponding field equations by assuming anisotropy. Section \ref{sec4} describes the physical requirements for a realistic compact star. In section \ref{sec5}, we presented particular cases of the model (by specifying metric potential) and obtained unknown parameters using matching conditions. Finally, in Section \ref{sec6}, we conclude our work by discussing the model's key findings.

\section{Einstein's field equations}
\label{sec2}
The interior of an anisotropic matter distribution in isotropic coordinates is described by the spherically symmetric spacetime metric
\begin{equation} \label{IMetric}
    ds^{2}=-A^2(r)dt^2+B^2(r)\left[dr^2+r^{2}\left(d\theta^{2}+\sin^{2}\theta d\phi^{2} \right)\right],
\end{equation}
where the metric potentials $A(r)$ and $B(r)$ are only functions of the radial coordinate r. The energy-momentum tensor for this matter distribution is given as
\begin{equation} \label{EMTensor}
    T_{ij}= diag(-\rho,p_{r},p_{t},p_{t}),
\end{equation}
where $\rho$ denotes energy density, and $p_{r}$ and $p_{t}$ represent radial pressure and tangential pressure, respectively, which are measured relative to the comoving fluid four-velocity
\begin{equation} \label{UFV}
    u^{a}=\frac 1{A} \delta^{a}.
\end{equation}
The Einstein's field equations for line element (\ref{IMetric}) and energy-momentum tensor (\ref{EMTensor}) reduces to the following system of equations in terms of physical parameters and metric function (using $G=c=1$)
\begin{equation} \label{Rho1}
    8\pi\rho=\frac{-1}{B^{2}}\Bigg[2\frac{B^{''}}{B}-\frac{{B^{'}}^{2}}{{B}^{2}}+\frac{4}{r}\frac{B^{'}}{B}\Bigg],
\end{equation}
\begin{equation} \label{Pr1}
    8\pi p_{r}=\frac{1}{B^2}\Bigg[\frac{{B^{'}}^{2}}{B^{2}}+2\frac{A^{'}}{A}\frac{B^{'}}{B}+\frac{2}{r}\Bigg(\frac{A^{'}}{A}+\frac{B^{'}}{B}\Bigg)\Bigg],
\end{equation}
and
\begin{equation} \label{Pt1}
    8\pi p_{t}=\frac{1}{B^2}\Bigg[\frac{A^{''}}{A}+\frac{B^{''}}{B}-\frac{{B^{'}}^{2}}{B^2}+\frac{1}{r}\Bigg(\frac{A^{'}}{A}+\frac{B^{'}}{B}\Bigg)\Bigg],
\end{equation}
where prime (') denotes differentiation with respect to the radial coordinate r.
Using equations (\ref{Pr1}) and (\ref{Pt1}), we get
\begin{equation} \label{Anisotorpy1}
    \Delta=p_{t}-p_{r}=\frac{1}{B^{2}}\Bigg[\frac{A^{''}}{A}+\frac{B^{''}}{B}-2\frac{{B^{''}}^{2}}{B^{2}}-2\frac{A^{'}}{A}\frac{B^{'}}{B}-\frac{1}{r}\Bigg(\frac{A^{'}}{A}+\frac{B^{'}}{B}\Bigg)\Bigg],
\end{equation}
here $\Delta=p_{t}-p_{r}$ is denoted as the anisotropic factor, and it measures the pressure anisotropy of the fluid, which is repulsive if $p_{t}>p_{r}$ and attractive if $p_{t}<p_{r}$.\\
On rearranging equation (\ref{Anisotorpy1}) we get
\begin{equation} \label{deq1}
    \frac{A^{''}}{A}-\left(2\frac{B^{'}}{B}+\frac{1}{r}\right)\frac{A^{'}}{A}=\Delta B^{2}-\frac{B^{''}}{B}+2\frac{{B^{'}}^{2}}{B^{2}}+\frac{1}{r} \frac{B^{'}}{B}.
\end{equation}

\section{Generating Model}
\label{sec3}
The Einstein's field equations for static spherically symmetric anisotropic fluid distributions are described by a system of three differential equations (\ref{Rho1}-\ref{Pt1}) with five unknowns $(\rho, p_r, p_t, A, B)$, and hence any two unknowns can be freely chosen to find exact solutions. To develop the model of a compact star, \cite{Bhar2015} choose density ($\rho$) along with radial pressure ($p_r$); \cite{SR2013} and \cite{BR2016} choose metric potential ($B^2$) together with radial pressure ($p_r$); \cite{MF2015} and \cite{TR2018} choose metric potential ($A^2$) with anisotropy ($\Delta$). A number of researchers [\cite{DELGADO2018}, \cite{SUDAN2017}, \cite{BOWERS1974}, \cite{COSENZA1981}] have examined the significance of pressure anisotropy in the modeling of compact objects in relation to the general theory of relativity. Our objective is to generate a singularity-free solution to the Einstein's field equations. In this model, we consider an appropriate form of the anisotropy ($\Delta$) and a particular form of the metric potential ($B^2$). 
We choose expression of an anisotropy ($\Delta$) as
\begin{equation} \label{anisotropy2}
    \Delta=\frac{B^{''}}{B^{3}}-2\frac{{B^{'}}^{2}}{B^{4}}-\frac{1}{r} \frac{B^{'}}{B^{3}}.
\end{equation}
Combining equation (\ref{anisotropy2}) in (\ref{deq1}),
\begin{equation} \label{deq2}
    \frac{A^{''}}{A}-\left(2\frac{B^{'}}{B}+\frac{1}{r}\right) \frac{A^{'}}{A}=0
\end{equation}
The solution of equation (\ref{deq2}) is given as
\begin{equation} \label{A}
    A=C \int{e^{\int{F\left(r\right)}dr}dr}+D,
\end{equation}
where
\begin{equation}
    F\left(r\right)=2\frac{B^{'}}{B}+\frac{1}{r},
\end{equation}
and C and D are constants of integration, which can be obtained by comparing the interior spacetime metric with the Schwarzschild exterior spacetime metric.\\
Therefore, spacetime metric (\ref{IMetric}) can be expressed as
\begin{equation}
\begin{split}
    ds^{2}=-{\left[C \int{e^{\int{F\left(r\right)}dr}}dr+D\right]}^{2} dt^{2}+ B^2(r)\\ \left[dr^2+r^{2}\left(d\theta^{2}+\sin^{2}\theta d\phi^{2} \right)\right].
    \end{split}
\end{equation}
Consequently, the expression of radial pressure and tangential pressure takes the form
\begin{equation} \label{Pr2}
    8\pi p_{r}=\frac{1}{B^2}\Bigg[\frac{{B^{'}}^{2}}{B^{2}}+2\frac{C e^{\int{F\left(r\right)}dr}}{C \int{ e^{\int{F\left(r\right)}dr}}dr+D} \left(\frac{B^{'}}{B}+\frac{1}{r}\right) +\frac{2}{r} \frac{B^{'}}{B} \Bigg],
\end{equation}
and
\begin{equation} \label{Pt2}
    8\pi p_{t}=\frac{1}{B^2}\Bigg[\frac{C e^{\int{F\left(r\right)}dr}}{C \int{ e^{\int{F\left(r\right)}dr}}dr+D} \left( 2\frac{B^{'}}{B}+\frac{2}{r}\right)+\frac{B^{''}}{B}-\frac{{B^{'}}^{2}}{B^{2}}+\frac{1}{r} \frac{B^{'}}{B} \Bigg],
\end{equation}

\section{Physical Requirements} \label{sec4}
To achieve a physically viable stellar configuration, the solutions of Einstein's field equations must satisfy the following conditions [\cite{Knusten1988}, \cite{Kuchowicz}, \cite{Buchdahl}].
\begin{itemize}
    \item[(i)] Regularity of the gravitational potentials at the origin:\\
The gravitational potentials $A(r)$ and $B(r)$ should be well-behaved at the center and regular throughout the interior of the star.
\item[(ii)] Positive definiteness of the density and pressure:\\
The energy density $(\rho)$, radial pressure $(p_r)$, and tangential pressure $(p_t)$ should be positively finite across the distribution.\\
i.e., $\frac{d\rho}{dr}<0$, $\frac{dp_{r}}{dr}<0$, $\frac{dp_{t}}{dr}<0$ for $0\leq r\leq R$.
\item[(iii)] Monotonic decrease of the density and pressure:\\
The density $(\rho)$, radial pressure $(p_r)$, and tangential pressure $(p_t)$ should decrease radially outward throughout the distribution.\\
i.e., $0<\frac{dp_{r}}{d\rho}<1$, $0<\frac{dp_{t}}{d\rho}<1$ for $0\leq r\leq R$.
\item[(iv)] Pressure Anisotropy:\\
The anisotropic factor $(\Delta=p_{t}-p_{r})$ should vanish at the center and increase monotonically outward.\\
i.e., $\Delta\left(r=0\right)=0$.
\item[(v)] Causality conditions:\\
The causality condition requires that the squared radial sound speed $\left({v^{2}_{r}}\right)$ and squared tangential sound speed $\left({v^{2}_{t}}\right)$ lie between 0 and 1.\\
i.e., $0<\frac{dp_{r}}{d\rho}<1$, $0<\frac{dp_{t}}{d\rho}<1$ for $0\leq r\leq R$.
\item[(vi)] Energy condition:\\
For an anisotropic stellar object, the strong energy condition should be satisfied.\\
i.e., $\rho-p_{r}-2p_{t}\geq 0$ for $0\leq r \leq R$.
\item[(vii)] Stability condition:\\
For stable stellar configurations, the adiabatic index $(\Gamma)$ should be greater than $\frac{4}{3}$.[\cite{Bondi1964}, \cite{Moustakidis}].\\
i.e., $\Gamma=\left(\frac{\rho+p_{r}}{p_{r}}\right)\frac{dp_{r}}{d\rho}>\frac{4}{3}$ for $0\leq r\leq R$.
\end{itemize}
These specific conditions help us in establishing the viability of an interior solution as a model for describing a compact star.
\section{Particular Cases}
\label{sec5}

It is well-known that obtaining solutions to the Einstein's field equations, which can describe a compact stellar structure meeting all physically acceptable conditions, is quite a challenge. One can make various choices for the metric potential $B\left(r\right)$ in order to solve the system of equations. It is worth noting that the model's gravitational and thermodynamic behaviour is determined by the metric potential $B\left(r\right)$. Therefore, the choice of metric potential $B\left(r\right)$ must meet all the criteria necessary for a realistic stellar model.

\subsection{Case I: If $\Delta=0$}
\begin{figure}
\centering
    \includegraphics[height=.2\textheight]{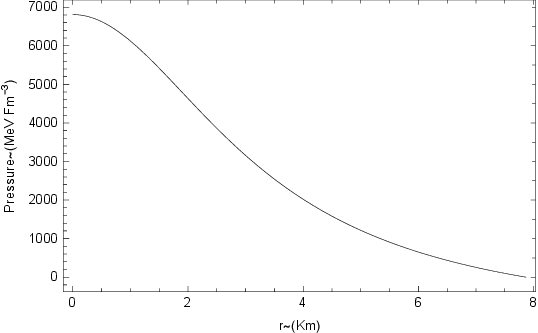}
    \caption{Variation of pressure $(p)$ against radial coordinate $(r)$ inside a stellar interior for a star 4U 1538-52 with $R = 7.87 km$, $a = 0.843$, and $b = 0.006$.}
    \label{fig:Figure. 1}
\end{figure}

\begin{figure}
\centering
    \includegraphics[height=.2\textheight]{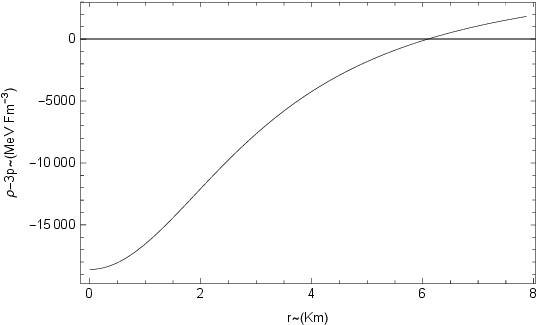}
    \caption{Strong energy condition $(\rho-3p)$ against radial coordinate $(r)$ inside a stellar interior for a star 4U 1538-52 with $R = 7.87 km$, $a = 0.843$, and $b = 0.006$.}
    \label{fig:Figure. 2}
\end{figure}

If pressure is isotropic in nature, then tangential and radial pressure are equal and measure of anisotorpy becomes zero, i.e.,
\begin{equation}
    \Delta=0,
\end{equation}
then equation (\ref{anisotropy2}) reduces to
\begin{equation}
    B^{''}-2\frac{{B^{'}}^{2}}{B}-\frac{1}{r} B^{'}=0.
\end{equation}
Solving the above differential equation
\begin{equation}
    B(r)=\frac{1}{a+br^{2}},
\end{equation}
where $a$ and $b$ are constants. Here, $B(0)$ = constant and $B^{'}(0) = 0$; it follows that the metric potential $B(r)$ is regular at the center and well-behaved throughout the star. With this choice of $B\left(r\right)$, metric potential $A\left(r\right)$ can be given as
\begin{equation}
    A(r)=\frac{-C}{2b\left(a+br^{2}\right)}+D,
\end{equation}
whee $C$ and $D$ are constants of integration.
Hence, spacetime metric (\ref{IMetric}) reduces to
\begin{equation} \label{Imetric1}
\begin{split}
    ds^{2}=-\left[\frac{-C}{2b\left(a+br^{2}\right)}+D\right]^{2} dt^{2}+\frac{1}{\left(a+br^{2}\right)^{2}}\\  \left[dr^{2}+r^{2}\left(d\theta^{2}+\sin^{2}\theta d\phi^{2} \right)\right].
    \end{split}
\end{equation}
To find the constants, we match the interior spacetime metric continuously to the exterior spacetime metric at the boundary r = R. The Schwarzschild exterior spacetime metric in isotropic coordinates is defined as
\begin{equation}\label{EMetric}
	ds^{2}=-\frac{\Big(1-\frac{M}{2r} \Big)^{2}}{\Big(1+\frac{M}{2r} \Big)^{2}}dt^{2}+\left(1+\frac{M}{2r} \right)^{4} \Bigg[dr^{2}
 + \\ r^{2}\left(d\theta^{2}+\sin^{2}\theta d\phi^{2} \right) \Bigg].
\end{equation}
A smooth matching of the metric potentials across the boundary is given by the first fundamental form, i.e.,
\begin{equation}
    A\left(R\right)=\frac{1-\frac{2M}{R}}{1+\frac{2M}{R}},
\end{equation}
\begin{equation}
    B\left(R\right)=\left(1+\frac{M}{2R}\right)^{2},
\end{equation}
and the second fundamental form suggests radial pressure vanishes at the surface of the star, i.e.,
\begin{equation}
    p_{r}\left(R\right)=0.
\end{equation}
The above boundary conditions help us to determine the constants as
\begin{equation}
    C=\frac{\left(2\sqrt{a+bR^{2}}-1\right) \left(2a^{2}b+2ab^{2}R^{2}\right)}{a-bR^{2}},
\end{equation}
\begin{equation}
    D=\frac{\left(2a-bR^{2}\right) \left(2\sqrt{a+bR^{2}}-1\right)}{a-bR^{2}}
\end{equation}
and
\begin{equation}
    M=2R\left(\frac{1}{\sqrt{a+bR^{2}}}-1\right).
\end{equation}
Therefore, metric potential $A\left(r\right)$ takes the form
\begin{equation}
    A\left(r\right)=\frac{\left(2\sqrt{a+bR^{2}}-1\right)R^{2}\left[\frac{a^{2}}{R^{2}}-b^{2}r^{2}-2ab\left(1-\frac{r^{2}}{R^{2}}\right)\right]}{\left(a+br^{2}\right) \left(a-bR^{2}\right)}.
\end{equation}
Consequently, the expressions for energy density $\left(\rho\right)$, radial pressure $\left(p_{r}\right)$ and tangential pressure $\left(p_{t}\right)$ can be given as
\begin{equation}
    \rho=12ab,
\end{equation}
\begin{equation}
    p_{r}=\frac{12a^{2}b^{2}\left(1-\frac{r^{2}}{R^{2}}\right)}{\frac{a^{2}}{R^{2}}-b^{2}r^{2}-2ab\left(1-\frac{r^{2}}{R^{2}}\right)},
\end{equation}
and
\begin{equation}
    p_{t}=\frac{12a^{2}b^{2}\left(1-\frac{r^{2}}{R^{2}}\right)}{\frac{a^{2}}{R^{2}}-b^{2}r^{2}-2ab\left(1-\frac{r^{2}}{R^{2}}\right)}.
\end{equation}
To demonstrate the radial dependency of our model's physical quantities, we considered the star \textbf{4U 1538-52}, whose estimated radius is $R = 7.87 km$ proposed by \cite{Gangopadhyay}. We choose $a = 0.843 km^{-2}$ and $b = 0.006 km^{-2}$. For these parameter values, we examined the behavior of energy density $(\rho)$ and pressure $(p=p_{r}=p_{t})$. Density $(\rho)$ is a constant in this case. Figure \ref{fig:Figure. 1} indicates that pressure $(p)$ reduces monotonically with increasing radial distance. Figure \ref{fig:Figure. 2} shows that the strong energy condition $(\rho-3p)$ does not hold since it is negative.

\subsection{Case II: If $B\left(r\right)=\frac{a}{\sqrt{1+br^{2}}}$}
A particular form of metric potential is chosen in order to solve the Einstein's field equations. We consider
\begin{equation} \label{B}
    B(r)=\frac{a}{\sqrt{1+br^{2}}},
\end{equation}
where a and b are constants. The gravitational potential $B\left(r\right)$ chosen above is well-behaved and finite at the origin since $B\left(0\right) =$ constant. Furthermore, $B^{'}\left(0\right)=0$ ensures that it is regular at the origin.
Substituting the value of $B\left(r\right)$ from equation (\ref{B}) in equation (\ref{A}), we get metric potential $A\left(r\right)$ as
\begin{equation} \label{A2}
    A=\frac{\left[-\sqrt{a}+2\left(1+bR^{2}\right)^\frac{1}{4}\right] \times C_{1}}{4\sqrt{a}},
\end{equation}
where
\begin{equation}
    C_{1} = 4+\left(2+bR^{2}\right)\log\left(1+br^{2}\right)
    -\left(2+bR^{2}\right)\log\left(1+bR^{2}\right).
\end{equation}
Using these metric potentials, the energy density, radial pressure, tangential pressure and anisotropy are obtained as
\begin{equation} \label{Rhoc1}
    \rho=\frac{b\left(6+br^{2}\right)}{a^{2}\left(1+br^{2}\right)},
\end{equation}
\begin{equation} \label{Prc1}
    p_{r}=\frac{b\left[4b\left(-r^{2}+R^{2}\right)-C_{2}+C_{3}\right]}{a^{2}\left(1+br^{2}\right)\left[4+C_{4}-C_{5}\right]},
\end{equation}
\normalsize
\begin{equation} \label{Ptc1}
    p_{t}=\frac{2b\left[2bR^{2}-C_{4}+C_{5}\right]}{a^{2}\left(1+br^{2}\right)\left[4+C_{4}-C_{5}\right]},
\end{equation}
and
\begin{equation}
    \Delta=\frac{b^{2}r^{2}}{a^{2}(1+br^{2})},
\end{equation}
where

\begin{equation}
C_{2}=\left(4+b^{2}r^{2}R^{2}+2b\left(r^{2}+R^{2}\right)\right)\log\left(1+br^{2}\right),
\end{equation}
\begin{equation}
C_{3}=\left(4+b^{2}r^{2}R^{2}+2b\left(r^{2}+R^{2}\right)\right)\log\left(1+bR^{2}\right),
\end{equation}
\begin{equation}
    C_{4}=\left(2+bR^{2}\right)\log\left(1+br^{2}\right),
\end{equation}
and
\begin{equation}
C_{5}=\left(2+bR^{2}\right)\log\left(1+bR^{2}\right).
\end{equation}\\
In order to examine the physical viability of the model, employ the radius of a particular star as an input parameter. To validate our model, we have considered the same star, \textbf{4U 1538-52}, having radius $R = 7.87 km$ with $a = 0.843 km^{-2}$ and $b = 0.006 km^{-2}$. Making use of such values for parameters, various physical quantities have been plotted graphically.

\begin{figure}
\centering
    \includegraphics[height=.2\textheight]{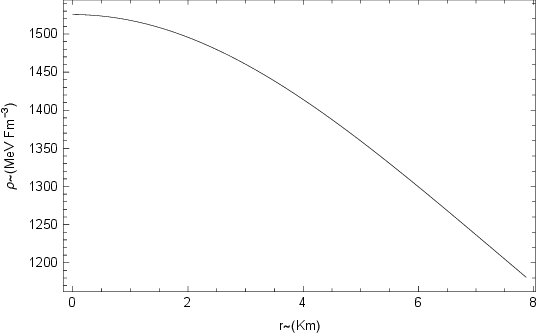}
    \caption{Variation of Density $(\rho)$ against radial coordinate $(r)$ inside a stellar interior for a star 4U 1538-52 with $R = 7.87 km$, $a = 0.843$, and $b = 0.006$.}
    \label{fig:Figure. 3}
\end{figure}

\begin{figure}
\centering
    \includegraphics[height=.2\textheight]{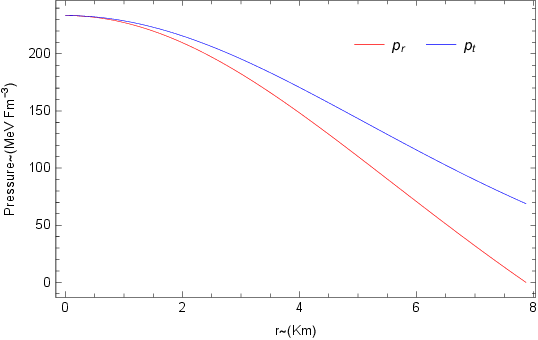}
    \caption{Variation of radial pressures ($p_{r}$ and $p_{t}$) against radial coordinate $(r)$ inside a stellar interior for a star 4U 1538-52 with $R = 7.87 km$, $a = 0.843$, and $b = 0.006$.}
    \label{fig:Figure. 4}
\end{figure}

\begin{figure}
\centering
    \includegraphics[height=.2\textheight]{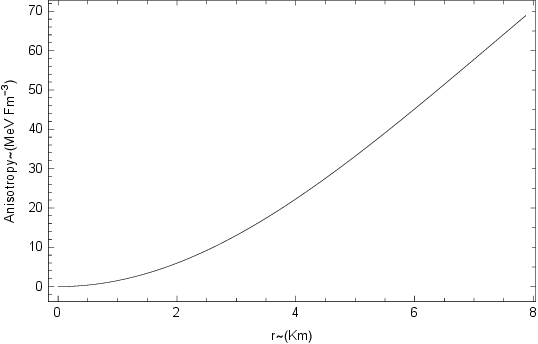}
    \caption{Measure of Anisotropy $(\Delta)$ against radial coordinate $(r)$ inside a stellar interior for a star 4U 1538-52 with $R = 7.87 km$, $a = 0.843$, and $b = 0.006$.}
    \label{fig:Figure. 5}
\end{figure}

\begin{figure}
\centering
    \includegraphics[height=.2\textheight]{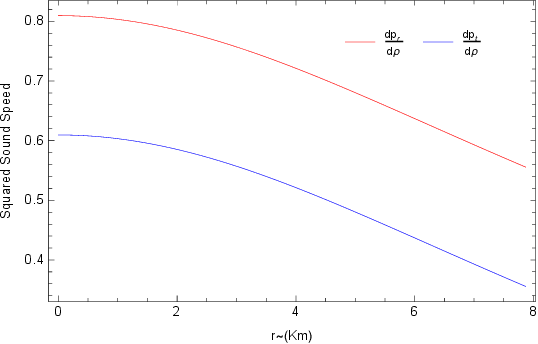}
    \caption{Squared Sound Speed against radial coordinate $(r)$ inside a stellar interior for a star 4U 1538-52 with $R = 7.87 km$, $a = 0.843$, and $b = 0.006$.}
    \label{fig:Figure. 6}
\end{figure}

\begin{figure}
\centering
    \includegraphics[height=.2\textheight]{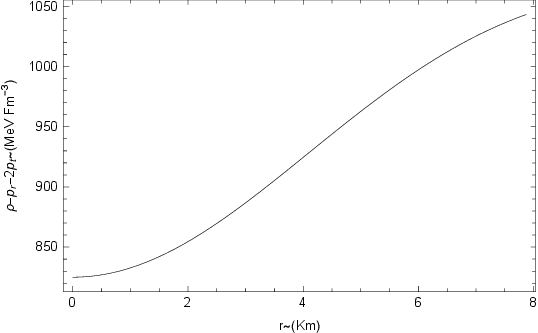}
    \caption{Strong Energy Condition ($\rho-p_{r}-2p_{t}$) against radial coordinate $(r)$ inside a stellar interior for a star 4U 1538-52 with $R = 7.87 km$, $a = 0.843$, and $b = 0.006$.}
    \label{fig:Figure. 7}
\end{figure}

\begin{figure}
\centering
    \includegraphics[height=.2\textheight]{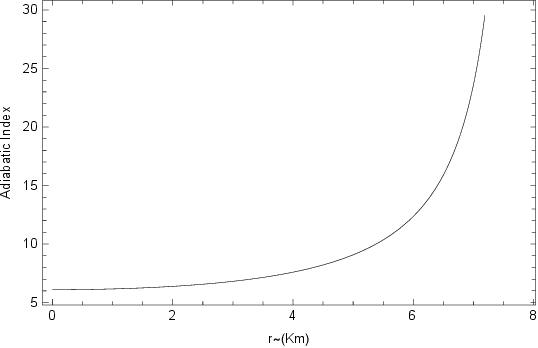}
    \caption{Adiabatic Index $(\Gamma)$ against radial coordinate $(r)$ inside a stellar interior for a star 4U 1538-52 with $R = 7.87 km$, $a = 0.843$, and $b = 0.006$.}
    \label{fig:Figure. 8}
\end{figure}

Figures \ref{fig:Figure. 3} and \ref{fig:Figure. 4} illustrate the energy density $(\rho)$, radial pressure $(p_{r})$, and tangential pressure $(p_{t})$, which are all positive and finite across the distribution. Figure \ref{fig:Figure. 3} illustrates that the energy density $(\rho)$ is maximum at the center of the star and decreases monotonically towards the surface. Figure \ref{fig:Figure. 4} depicts the variation of radial and tangential pressures against the radial coordinate $(r)$, which also decreases radially outward from its maximum value at the center of the star, and the radial pressure vanishes at the surface, whereas the tangential pressure remains non-zero at the surface. Figure \ref{fig:Figure. 5} depicts the measure of anisotropy $(\Delta)$, which is zero at the center and reaches its maximum at the boundary. A positive anisotropic factor $(\Delta>0)$, or repulsive nature, enhances system stability and allows for the construction of more compact structures [\cite{GM1994}]. Figure \ref{fig:Figure. 6} displays the squared radial and tangential sound speeds. Speeds remain within the acceptable range throughout the stellar interior. This feature ensures that the causality condition is not violated, indicating that the model is potentially stable throughout the stellar interior [\cite{H1992}, \cite{Abreu}]. Figure \ref{fig:Figure. 7} depicts the strong energy condition, which is positive throughout the stellar interior and essential for a physically feasible stellar model. Figure \ref{fig:Figure. 8} shows a graph of the adiabatic index $(\Gamma)$, indicating that $\Gamma$ exceeds $\frac{4}{3}$ throughout the stellar configuration. As a result, one could conclude that the model is stable [\cite{Ponce}]. The detailed study of the above model can be found in the model developed by \cite{Suthar2024}.

\subsection{Case III: If $B\left(r\right)=\frac{1}{\left(1+r^{2}\right)^{\frac{1}{7}}}$}
We consider
\begin{equation}
    B\left(r\right)=\frac{1}{\left(1+r^{2}\right)^\frac{1}{7}}.
\end{equation}
Since $B\left(0\right)=$ constant and $B^{'}\left(0\right)=0$, the above metric potential is regular at the origin and well behaved in the interior of the sphere. With this choice of $B\left(r\right)$, the other metric potential is given as
\footnotesize
\begin{equation}
\begin{split}
A\left(r\right)=\left[19R^{4}+R^{2}\left(6\left(1+r^{2}\right)^{\frac{5}{7}}\left(1+R^{2}\right)^{\frac{2}{7}}+47\right)+7\left(\left(1+r^{2}\right)^{\frac{5}{7}}\left(1+R^{2}\right)^{\frac{2}{7}}+4\right)\right]\\ * \frac{\left[2\left(1+R^{2}\right)^{\frac{1}{14}}-1\right]}{5\left(5R^{4}+12R^{2}+7\right)} 
\end{split}
\end{equation}
\normalsize
Subsequently, the expressions for energy density, radial pressure and tangential pressure can be given as
\begin{equation}
    \rho=\frac{12\left(7+2r^{2}\right)}{49\left(1+r^{2}\right)^{\frac{12}{7}}},
\end{equation}
\begin{equation}
\begin{split}
    p_{r}=\frac{4r^{2}\left[114R^{2}C_{6}+133 C_{6}-6\left(1+R^{2}\right)\left(28+19R^{2}\right)\right]}{49\left(1+r^{2}\right)^{\frac{12}{7}}\left[19R^{4}+R^{2}\left(6 C_{6}+47\right)+7\left(C_{6}+4\right)\right]} +\\ \frac{28\left[-19R^{4}+R^{2}\left(24C_{6}-47\right)+28\left(C_{6}-1\right)\right]}{49\left(1+r^{2}\right)^{\frac{12}{7}}\left[19R^{4}+R^{2}\left(6 C_{6}+47\right)+7\left(C_{6}+4\right)\right]},
    \end{split}
\end{equation}

\begin{figure}
\centering
    \includegraphics[height=.2\textheight]{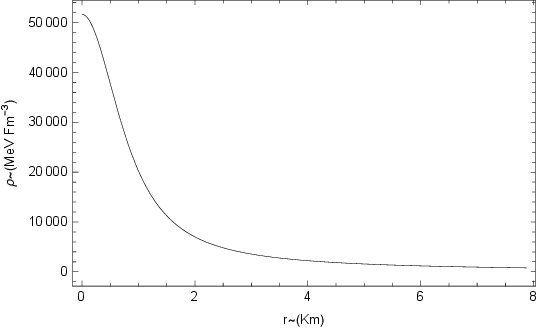}
    \caption{Variation of energy density $(\rho)$ against radial coordinate $(r)$ inside a stellar interior for a star 4U 1538-52 with $R = 7.87 km$}
    \label{fig:Figure. 9}
\end{figure}

\begin{figure}
\centering
    \includegraphics[height=.2\textheight]{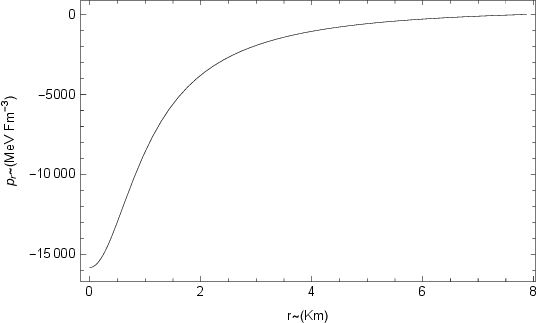}
    \caption{Variation of radial pressure $(p_{r})$ against radial coordinate $(r)$ inside a stellar interior for a star 4U 1538-52 with $R = 7.87 km$}
    \label{fig:Figure. 10}
\end{figure}

\normalsize
and
\small
\begin{equation}
p_{t}=\frac{-532R^{4}+4R^{2}\left[150r^{2}C_{6}+168C_{6}-329\right]+28\left[25r^{2}C_{6}+28C_{6}-28\right]}{49\left(1+r^{2}\right)^{\frac{12}{7}}\left[19R^{4}+R^{2}\left(6 C_{6}+47\right)+7\left(C_{6}+4\right)\right]},
\end{equation}
\normalsize
where
\begin{equation}
    C_{6}=\left(1+r^{2}\right)^{\frac{5}{7}} \left(1+R^{2}\right)^{\frac{2}{7}}.
\end{equation}
To conduct an in-depth evaluation of different physical conditions, we have considered the same star, \textbf{4U 1538-52}, which has a radius $R = 7.87 k$m. Using this value of radius as an input parameter, we have presented the behavior of energy density and radial pressure in Figures \ref{fig:Figure. 9} and \ref{fig:Figure. 10}, respectively. Figure \ref{fig:Figure. 9} depicts that the energy density is monotonically decreasing throughout the distribution. Figure \ref{fig:Figure. 10} shows the radial pressure, which is negative within the stellar interior. However, the models generated in this manner are not always physically plausible.



\section{Summary and conclusions}
\label{sec6}
In this work, we obtained an exact analytic solution to the Einstein's field equations for a spherically symmetric matter distribution in isotropic coordinates. To develop a physically acceptable stellar model, many authors imposed an equation of state, which is a relationship between pressure and density. Since the equation of state might not be known at very high densities, we developed the model where the equation of state is not exactly known. The solution generated by assuming a generalized physically reasonable metric potential $B(r)$ and a particular form of anisotropy turns out to be non-singular, and regular, and it could be utilized to describe relativistic compact stars. We did not choose a specific form of $B(r)$, and we defined the anisotropy using $B(r)$ and its derivative with respect to r. As a result, we can implicitly conclude that there is only one generator, $B(r)$. We discussed one physically reasonable model. Anisotropy becomes zero for a particular choice of $B(r)$, resulting in the isotropic solution stated as one of the particular cases. The new class of generalized solutions reduces to the \cite{Suthar2024} model for specific choices of the metric potential $B(r)=\frac{a}{\sqrt{1+br^{2}}}$. The model parameters were found by assuming that the interior and exterior metrics at the boundary of the star are continuous and that the radial pressure at the boundary vanishes. Hence, one can obtain all possible solutions by choosing different metric potentials $B(r)$. However, the models developed via this approach are not necessarily physically plausible. The model proposed here can be considerably studied to accommodate astrophysical objects.

\section*{Acknowledgements}
BSR and BS gratefully acknowledge support from the Inter-University Centre for Astronomy and Astrophysics (IUCAA), Pune, India, for the hospitality and facility provided to them where part of the work was carried out.




\bibliographystyle{elsarticle-harv} 
\bibliography{manuscript} 






\end{document}